\documentclass[12pt]{article}
\usepackage{graphicx}
\usepackage{fancyhdr}
\usepackage{lastpage}

\begin{document}
\begin{center}
\begin{bfseries}

\LARGE{Influence of the topology on the power flux of the Italian high-voltage electrical network} \\
\end{bfseries}
\vspace {1 cm}
V.Rosato$^{1,2}$, L. Issacharoff$^{1}$ , G. Gianese$^{2}$ and S.Bologna$^{1}$ \\

 \vspace {1 cm}

$^{1}$ ENEA, Casaccia Research Center, Computing and Modelling Unit, Via Anguillarese 301, 00060 S.Maria di Galeria, Roma (Italy)\\
$^{2}$Ylichron Srl.,ENEA, Casaccia Research Center, Via Anguillarese 301, 00060 S.Maria di Galeria, Roma (Italy)\\

\end{center}

\begin{abstract}
A model of the Italian 380 kV electrical transmission network has been analyzed under the topological and the functional viewpoints. The DC power flow model used to evaluate the power flux has been solved on the basis of input conditions (injected power - extracted power, line's reactances and the maximum flux capacity of each line) taken from real data. The vulnerability of the network under load conditions has been estimated by evaluating the power flux redistribution along the lines subsequent to line's removal. When the perturbed network cannot sustain a given input--output demand, the maximum power sustainable by the network has been evaluated to optimize the \texttt{Quality of Service}, defined as the difference between the expected and the effective dispatched power. The functional relevance of the different lines of the network has been classified according to the amount of power that the network must reduce, to keep alive, upon their removal. Results show that topological and functional relevances are related to different lines; lines having a strong topological relevance may have a bare relevance in the flow distribution and thus their removal does not affect the functioning of the network.

\end{abstract}

\section{Introduction}
The high-voltage electrical transmission network is at the top of a hierarchical view of electrical Critical Infrastructures (CI); its incorrect functioning has a strong repercussion on both the electrical  transmission and distribution processes. Its functioning, moreover, strongly influences that of many other CIs which are mutually \texttt{interdependent}(see, in this respect, \cite{CNIP06} and the assessment of the effects of the blackout experienced by Italy on September 2003 \footnote{Interim Report of the Investigation Committee on the 28th September 2003 blackout in Italy (http://www.ucte.org/publications/library). UCTE is the "Union for the Co-ordination of Transmission of Electricity" ; it represents the association of transmission system operators in continental Europe, providing a reliable market base by efficient and secure electric "power highways".}).

In the present work, we study the properties of the italian high-voltage (380 kV) electrical transmission network (HVIET hereafter), to estimate the vulnerability of the network on the basis of the analysis of its topological properties and of the results of a model of power transport. We have also attempted to establish some correlation between  the $topological$ properties of the different parts of the network and their functional relevance.

Several works (see, among others, \cite{Albert1,crucitti,EPSR}) have pointed out the existence of some relation between topology and vulnerability. Topological analysis might be used to predict points of $structural$ vulnerability (i.e. indicating the links whose failure induce a severe $structural$ damage like, $e$. $g$., the disconnection of some node or the increase of the network's diameter or the average internode's distance). In the present work, we wish to investigate the vulnerability issue also from the viewpoint of the network's $functioning$. The electrical power flow establishes along the different lines, depending on topology, on the position in the networks of sources and loads, on the line's characteristics etc. It could happen that lines having topological relevance have, in turn, a bare relevance on the flow distribution. It is therefore important to associate to a vulnerability study based on the simple $topological$  analysis of its components, the analysis of the impact that faults (i.e. arcs and/or node's removal) have on the network's functionality.

\section{Model and computational method}
We have analyzed data relative to the Italian high-voltage (380 kV) electrical transmission network (HVIET). This study follows previous works performed to unveil, from topological analysis, a number of features of the HVIET \cite {crucitti,EPSR}. The present work, however, aims at introducing a vulnerability analysis based on a $dynamical$ model of the network, by reproducing the power flow conditions and by evaluating the impact on these conditions produced by the introduction of faults in the form of line's removal.

New network's data have been inferred from the analysis of the public documentation. They slightly differ from those recently reported in a previous work \cite{EPSR} as they account for the opening of new tracts of the network. HVIET can be represented by an undirected graph of $N$ nodes and $E$ lines ($N=310$ and $E=361$). Data allow to qualify  each node as $source$ $S$ (where part of the power is inserted in the network), $load$ $L$ (where part of the power is extracted from the network) or  $junction$ $J$ (which is neither a $S$ nor a $L$ node).  There are $S=97$ source nodes, $L= 113$ load nodes and $J= 100$ junction nodes.
The topology of the HVIET is reported in Fig.1. Some lines (14 out of 361) are constituted by $double$ lines which are not displayed in the graph of fig.1.
Points of cut of the network (which is connected with other european networks) have been substituted with "fictitious" source nodes where the same amount of electrical power received by foreigner countries is pumped into the network.

Several topological properties have been analysed: first of all, the distribution $P(k)$ of the node's $degree$ $k$ which allows to "classify" its topology.  The network has a limited number of hubs, whose maximum degree is $k_{max}=11$.  $P(k)$ and the cumulative degree distribution $P(k>K)$ are both likely to be fit to an exponential (single--scale network \cite{Albert1,crucitti,amaral1}). The latter distribution can be fitted to $P(k>K)\sim e^{-0.55K}$ in agreement with previous findings for the North-american power grid  \cite{Albert1}.
A further property measured on the HVIET network is the average $clustering$ coefficient $C$ \cite{crucitti} which results to be as small as $C= 2.06 \cdot 10^{-2}$.

 We have also evaluated other relevant topological properties of nodes and lines: the $betweenneess$ and the $information$ centralities.

 $Betwenneess$ centrality is defined for nodes and lines, $b_i$ and $b_{ij}$ respectively, as in \cite{crucitti}, by measuring the "relevance" of a node (link) for the connection of all other nodes.

\begin{equation}
b_i = \frac{1}{(N-1)(N-2)}\sum_{j,k \neq i} \frac{n_{jk}(i)}{n_{jk}}
\end{equation}

\begin{equation}
b_{ij} = \frac{1}{N_p}\sum_{k,l} \frac{n_{kl}(ij)}{n_{kl}}
\end{equation}

where ${n_{jk}(i)}$ is the number of shortest paths connecting nodes $j$ and $k$ which pass through node $i$; ${n_{kl}(ij)}$ is the number of shortest paths connecting nodes $k$ and $l$ which make use of the link between nodes $i$ and $j$ (if any) and $N_p$ the total number of paths on which the sum is evaluated.

 A further measured quantity has been the $information$ centrality of the nodes (introduced by \cite{latora2}), which relates the importance of a node to the ability of the network to respond to its removal.
If one defines the $efficiency$ $E[G]$ of a graph as
\begin{equation}
E[G] = \frac{1}{N(N-1)}\sum_{i,j\in G} \frac{1}{d_{ij}}
\end{equation}
where $d_{ij}$ is the shortest--path distance between nodes $i$ and $j$, then the $information$ centrality  $I_i$ of node $i$ is
\begin{equation}
I_i = \frac{\Delta E}{E} = \frac{E[G]-E[G']}{E[G]}
\end{equation}
where $E[G]$ and $E[G']$ are the $efficiency$ of the unperturbed network and that of the network after the removal of node $i$, respectively. Analoguosly to the $betwennees$, also the $information$ centrality can be defined for lines, giving origin to the quantity $I_{ij}$.

Table 1 provides the ranking of the nodes with respect to the values of the $degree$ $k$, the $betweenness$ centrality $b_{i}$ and the $information$ centrality $I_{i}$, while Table 2 gives the ranks of the lines with respect to $b_{ij}$ and $I_{ij}$.

Each graph can be represented by an $Adjacency$ matrix $\bf A$ whose elements $A_{ij}=1$ if nodes $i$ and $j$ are connected (there is a link in between). A further insight on the network's topology can be obtained by performing the spectral analysis of its Laplacian matrix $\bf L$ defined as
\begin{equation}
L_{ij} =  \left \{
    \begin{array}{rl}
    \sum_{k=1}^{N} A_{ik} & \mbox{if } i = j \\
    - A_{ij}   & \mbox{if } i \ne j
    \end{array}
     \right.
\end{equation}
The analysis of the sign of the components of the eigenvector associated to the first non--vanishing eigenvalue of the Laplacian allows to optimally bisecate the network. As $\bf L$ is symmetric, the first eigenvalue is always vanishing. The $n$ components of the eigenvector $\textbf{v}^L_{2} = (v_{1}, v_{2}, ..., v_{n})$ associated to the second eigenvalue solve the one--dimensional \textit{quadratic placement} problem of minimizing the function
\begin{equation}
z=\frac{1}{2} \sum_{i=1}^{n} \sum_{j=1}^{n} (v_{i}-v_{j})^{2}A_{ij}
\end{equation}
The vector is subject to the constraint $\left| \textbf{v} \right| = (\textbf{v}^{T}\textbf{v})^{\frac{1}{2}} = 1$ \cite{Hagen}. The process allows to partition the graph $G = (N, E)$ into two disjoint subsets $U$ and $W$ such that $L_{UW}/(N_{U} \cdot N_{W})$ is minimized. $L_{UW}$ is the number of links to be removed and $N_{U}, N_{W}$ are the number of nodes in the two resulting subnetworks. It comes clear that this procedure allows the "optimal" bisection of the graph, i.e. it forms the closest possible subnetworks $U$ and $V$ with the minimum amount of broken links $L_{UW}$ \cite{mohar,pothen}.
This allows to define a $critical$ $section$, an ideal cut line which bisecates the network into the two largest connected subnetworks with the minimum number of broken links.

By applying of the min--cut method to HVIET  \cite{EPSR} the network is diveded into two, connected, parts HVIET$_U$ and HVIET$_W$, the first formed by a number of nodes N(HVIET$_U$)=115, the second by N(HVIET$_W$)=195. The two parts are separated by $only$ six links (see fig.1).
The ideal line, joining the location of the removed links, could be called \texttt{first critical section}. Table 3 reports the links involved in the first critical section of the network. This is a major outcome of the spectral analysys providing a way to locate the critical vulnerability lines of the network. This procedure could be, in fact, iterated on the resulting components of the graphs, by creating \texttt{critical sections} of higher orders. The definition of critical sections of higher order could be a valuable solution for the problem of \texttt{islanding}, a procedure often used to isolate regions of the network to avoid cascade effects \cite{islanding}.

There are a few nodes with critical topological properties: nodes $158$, $183$, $153$, $168$ and $154$ have highest values of $b_i$ and $I_i$. These nodes are also important as starting/ending points of lines of topological relevance (high $b_{ij}$ and $I_{ij}$).
A few of these nodes ( $153$, $183$ and $154$) are also involved in the first critical section. All these nodes could be thus ascribed to a set of node whose failure produces a severe "topological" damage.

We will now investigate if the electrical flow makes use of  these nodes or if the interplay between topological and electrical properties of the different lines modifies this scenario.

\section{The DC power flow model}
We now introduce a flow model for the electrical power which will allow to analyse the "dynamical vulnerability" of the network.

We will use a simplified transport model as we wish to evaluate the effect of the network topology on the steady--state power flow rather than on transitory regimes. We will firstly evaluate the power flow distribution in the unperturbed network, resulting from a specific "power input"--"power output" condition, chosen to be representative of a typical power requirement that HVIET must daily sustain. Then, we will perturb the network, by removing links, and we will detect if the "damaged" network is still able to produce a correct response to the input--output demand, within the imposed physical constraints.

 We have modeled the transport of the electrical flow in the network by using a DC power flow model.
The DC power flow equations \cite{DC} provide a linear relationship between the active power flowing through the lines and the power input into the nodes. They can be formulated  as follows:
\begin{equation}
\label{Pkm}
F_{km}=\frac{\theta_k-\theta_m}{x_{km}}
\end{equation}
where $x_{km}$ is the series reactance of the line connecting nodes $k$ and $m$, $F_{km}$ is the active power flow on this line and $\theta_k,\theta_m$ are the voltage phase of the $k$'th and $m$'th nodes.
Summing on all branches connected to the node $k$, the power flow of that node $P_k$ is
\begin{equation}
P_k=\sum_m{F_{km}}=\theta_k\sum_m{x_{km}^{-1}} - \sum_m{\frac{\theta_m}{x_{km}}}
\end{equation}
This can be written in the matrix form
\begin{equation}
\label{dc_equ}
{\bf P}={\bf B}{\bf \theta}
\end{equation}
where $B_{km}=-1/x_{km}$ (for $k \neq m$) and $B_{kk}=\sum_l{1/x_{kl}}$. For the calculation of $B$, double lines were regarded as a single line and its reactance adjusted accordingly. ${\bf B}$ is a $N\times N$ matrix. Its rank is, however, $N-1$ since the network must comply with the conservation condition $ \sum_{i=1}^{N} P_i = 0$.
To solve the system, an equation is removed and the associated link is chosen in a way to introduce a reference node whose phase angle is arbitrarily set to $\theta=0$.
For a given input vector $\bf P^{(0)}$ and a given set of line's reactance $x_{km}$, the linear system in eq. (\ref{dc_equ}) can be solved and a solution is found in terms of $\theta_{ij}$ and $F_{ij}$. Two constraints must be imposed to ensure the physical correctness of the solution. These originate from the fact that the DC power flow method results from the elimination of the imaginary part of the current equations, under the hypothesis that power phase angles are small. It should thus result, $\forall {(k,m)}$, that
\begin{enumerate}
\item{$\theta_{km} <30$ degrees}
\item{$\mid F_{km}\mid <F_{km}^{max}$ (where$F_{km}^{max}$ is some specified limiting power flux which can be sustained by the line between nodes $k$ and $m$).}
\end{enumerate}
If constraint (1) is not fulfilled, the inductive part of the electrical flux cannot be disregarded and eq.(\ref{dc_equ}) does not hold. Constraint (2) is a technological threshold, relating to the specific line's impedance. A too large flux  produces an unendurable heat which is normally prevented, in real transmission networks, by $ad$ $hoc$ elements which disconnect the line.

Data needed to solve the DC power flow system are thus:
\begin{itemize}
\item{(a) the vector of input-output demand $P^{(0)}_i$ ($i=1,N$). Data used for the simulation represents a typical snapshot of the vector (injected power--extracted power) experienced daily in the HVIET; they sum up to a total power injected in the line of the order $P_{tot}=31.8 GW$ which represents an average power demand in a spring--time working day in Italy- May 2005 data). Source nodes are characterized by negative $P_i$ values, junction by vanishing values, loads by positive values;}
 \item{(b) the line's reactance $x_{ij}$;}
 \item{(c) the values of the maximum sustainable power flux $F_{ij}^{max}$ for each line.}
\end{itemize}

The solution of the DC power flow system allows to evaluate: (1) the power flux $F_{ij}$ along the lines resulting from the given input vector  $\bf P^{(0)}$; (2) the phase angles $\theta_{ij}$. These results will constitute the "normal" response of the network to the input conditions $\bf P^{(0)}$.
Lines with the highest flux value are reported in Table 4: second column contains the line's initial and ending nodes, third column contains the power flux value.

\section{Perturbations and dynamical vulnerability}
Structural perturbations are imposed by removing an increasing number of lines, and measuring the resulting distribution of the power flux as a function of the perturbation strength $\xi$, defined as the number of simultaneously--removed lines. Disregarding the lines whose removal produces the disconnection of one (or more) node (in these cases, treated in \cite{EPSR}, the DC power flow equations cannot be solved), the more vulnerable parts of the networks will be those lines whose removal will perturb the network at the point that it will not be able anymore to fulfill the dispatching of the required power vector $\bf P^{(0)}$ (i.e. within the imposed constraints (1) and (2) above).

Starting from the input vector $\bf P^{(0)}$, we have carried out two types of simulations.

In the first type, we removed an increasing number of lines: if the power flux established in the perturbed network is not consistent with the required constraints, the system is defined as $unsatisfied$ and its level of \texttt{Operability} (defined through a $Quality$ $of$ $Service$ ($QoS$) function) is set to zero. For a given perturbation strength, there will be cases in which the removal of specific lines produces a vanishing $QoS$, others in which, in turn, it produces a power flow redistribution still fulfilling the imposed contraints (in such a case, $QoS=1$). In the cases where a part of the network (one single node or an entire subnetwork) is disconnected, the DC power flux equations cannot be solved and the dispatching problem cannot be fulfilled; also in this case $QoS=0$. The value of the $QoS$ associated to a given perturbation strength will result by the average over the different values of the $QoS$ issued upon the different choices of lines removal ($<QoS>$ in fig.2 left).
 We have firstly separately  removed each line ($\xi=1$). Then, we have removed all pairs ($\xi=2$), triplets ($\xi=3$) and quadruplets ($\xi=4$) of lines. As, in the cases $\xi>2$ an exaustive evaluation of all possible line's combinations cannot be attempted, we have evaluated the average over a large number of different cases (of the order of $10^4$) for $\xi=3$ and $\xi=4$.

In the second type of simulations, we repeat the same as in the first type with the difference that, when the perturbation is such to inhibit the system's satisfaction, we implement an optimization strategy determining the optimal variation of the input vector $\bf P$ able to re--establish a correct power flow (i.e. within the imposed constraint). We called this procedure "re--dispatching". The "optimal" value is thus searched by reducing the input power (and the corresponding output power) in a way to maximize the function $QoS$

\begin{equation}
QoS = 1 - \frac{\sum _{i\in {loads}}[P^{(0)}_i - P_i]}{\sum_{i\in {loads}} P^{(0)}_i}
\end{equation}

where the value of $P_i$ is the input vector resulting by optimization. Re--dispatching allows the fulfilment of the electrical problem, although with some degradation, measured by the resulting value of the $QoS$.

The general procedure of the re--dispatching is the following. If we define as $\tilde{S}$ and $\tilde{L}$ the set of all sources and loads, respectively, and $\tilde{NL}$ as the number of loads in the $\tilde{L}$ list, it must:
\begin{enumerate}
\item give all nodes zero value ($P_i=0 ~~ \forall ~ i$)
\item choose a source node $i\in \tilde{S}$.
\item increase $P_i$ such that $P_i=P_i+\epsilon$
\item increase value of all loads (still below required consumption) such that
$\forall j\in \tilde{L} ~~~ P_j=P_j-\epsilon/\tilde{NL}$ (loads are measured in negative power units).
\item solve DC flow equations and, if constraints (1) and (2) are fulfilled, accept the changes and choose a different source node ($i=i+1$). else,
if constraints are not kept, change again the values of loads and sources to what it was before step 3.
\item go to step 2
\end{enumerate}


If, at step 5, changes are accepted, sources and loads which have reached the required value (defined in the formulation of the problem by $\bf P^{(0)}$) are taken out of the list. The procedure will continue until both sets are empty or until no more changes can be made. Fig.2 contains also the results of the behavior of $<QoS>$ in the re-dispatching case, for $\xi=1-4$. It is clear that, in absence of re-dispatching actions, as soon as $\xi > 1$, the $<QoS>$ value starts decreasing. The electrical network is, in fact, designed to run properly, in practically all load conditions, for $\xi=1$, unless the cases in which nodes are disconnected with a single line removal (this happens to nodes with $k=1$, called $leaves$). If one operates the (hypothetical) re-dispatching procedure which has been proposed in the simulations, the system can be brought to resist to severe perturbations with a resulting quite small degradation. It has to be considered, however, that the type of re-dispatching prefigured in the present work cannot be easily produced in a limited amount of time and, in some cases, it also requires a huge technical effort. Simulations results have been produced to show how a rapid and wise action on the power sources and loads could significantly reduce the impact of severe faults on the network (see, e.g., the resulting value of the $<QoS>$ upon re-dispatching for $\xi=4$).

The results of the re--dispatching strategy can be also used to establish a hierarchy of lines, on the basis of the power reduction associated to their removal. In other words, when $\xi=1$, each line is removed and the corresponding re--dispatching procedure is performed if the removal of that line produces the system's inoperability. If one relates the specific line with the associated value of $\Delta P = \sum _{i\in {loads}}[P_i(L_0) - P_i]$ one obtains the graph reported in fig.3. The indication of the electrical lines whose removal requires the highest amount of redispatched power to restore electrical flux is also reported in Table 5. Data show that there are links whose removal introduces a severe perturbation: although the network is still able to sustain a power flow, its global amount should be significantly reduced. There are a few lines whose removal reduces the overall network's capacity of an amount around 1500 MW.

The obtained results show that only a small number of nodes having a "functional" relevance for the network can be discovered through the topology analysis of its graph. The inspection of Tables 4 and 5 shows that critical lines for electrical transport are related to nodes $214$, $184$, $117$, $190$, $127$, $130$ etc.; this apparently does not appear in the first ranks of nodes (and lines) with high topological relevance. 

A more detailed quantitative assessment of this point can be obtained by plotting, for each line of the network, the relative value of the $information$ centrality $I_{ij}$ and that of the associated $\Delta P$ (fig.4). This figure compares a topological relevant quantity (the line $information$ centrality), related to the "structural" damage produced by the removal of a line, with a functional relevant quantity (the value of $\Delta P$). It comes clear that, disregarding the lines whose removal produces an irreparable damage (i.e. one or more nodes are disconnected and DC power flow equations cannot be solved anymore; in that case the value of $\Delta P$ is arbitrarily set to $\infty$), high values of $\Delta P$ are not necessarily associated to large values of $I_{ij}$. This observation validates the assumption that "structural" and "functional" vulnerability are associated with different sets of nodes and lines.

\section {Conclusions}
This work has shown how it is possible to set up a functional model of an electrical network enabling to relate the network's topology to its level of function. A simple DC power flow method has been used to evaluate the "efficiency" of the network to sustain the flux of a given amount of electrical power, deduced by real data relative to the italian dispatching of high--voltage electrical power (380 kV). Topological analysis of the network has put in evidence a number of lines which are related to the network's structural vulnerability. These lines have been associated either to links with high centrality properties, either to links belonging to the first critical section resulting from the spectral analysis of the Laplacian matrix associated to the network's graph. The DC power flow simulation has allowed to establish a hierarchy of network's lines to which is associated a high functional vulnerability (the removal of these lines can be counterbalanced only by considerably reducing the power flow level that the network can sustain).
Comparing the two sets of data (lines of high structural vulnerability and lines of high functional vulnerability) we have put in evidence that the electrical flow simulations qualify as relevant lines which are not those identified as lines of maximal structural relevance. For this reason, it should be remarked that topological analysis and the simulation of "functional" models (such as the DC power flow model, for the case of electrical networks) provide $complementary$ informations; $both$ their results must be used to correctly predict the vulnerability properties of complex networks.

\section{Acknowledgments}
This work has been performed under the framework of the EU project "IRRIIS" (Integrated Risk Reduction of Information-based Infrastructure Systems) under the the IST programme of the Sixth Framework Programme (FP6-2005-IST-4). The authors acknowledge discussions and suggestions by C. Balducelli and M. Minichino (ENEA). M.Delfanti and C.Bovo (Milan Polytechnic) are kindly acknowledged for their useful suggestions on the use of the DC power flow .

\newpage

\begin{table}
\caption{Rank of nodes on the basis of their $degree$ $k$ (second column); $betweenneess$ centrality  $b_i$ (third column) and $information$ centrality $I_i$ (fourth column).}
\begin{center}
\begin{tabular}{|c|c|c|c|}
\hline
rank& highest $k$ ($k$) &highest $b_i$ & highest $I_{i}$ \\
\hline
1&178(11)&183 &201 \\
2&158(11)&168 &183 \\
3&201(10)&153 &158 \\
4&211(8) &207 &214 \\
5&200(8) &184 &103 \\
6&124(8) &154 &152 \\
\hline
\end{tabular}
\end{center}
\end{table}

\newpage

\begin{table}
\caption{Rank of links on the basis of their  $betweenneess$ centrality  $b_{ij}$ (second column) and $information$ centrality $I_{ij}$ (third column). Links are displayed by the initial and ending node's numbers.}
\begin{center}
\begin{tabular}{|c|c|c|}
\hline
rank& highest $b_{ij}$ & highest $I_{ij}$ \\
\hline
1& 153--168& 183-207 \\
2& 153--183& 153-183\\
3& 183--207& 100-191\\
4& 170--204& 183-233\\
5& 109--158& 178-263\\
6& 161--183& 192-226\\
\hline
\end{tabular}
\end{center}
\end{table}

\newpage

\begin{table}

\caption{\label{}List of links belonging to the first critical section (see text).}

\vspace{0,5cm}
\begin{center}
\begin{tabular}{|c|c|c|c|c|c|}
\hline
137-154 & 145-143& 145-169& 153-183& 191-169& 212-154\\
\hline
\end{tabular}
\end{center}
\end{table}

\newpage

\begin{table}
\caption{Rank of lines on the basis of the flux they sustain in normal (i.e. under the $P^{(0)}$ power vector) operating conditions.}
\begin{center}
\begin{tabular}{|c|c|c|}
\hline
rank& line & flux (MW) \\
\hline
1&214--184& 1324 \\
2&117--190&  1295  \\
3&117--201&  1270\\
4&120--190& 1087   \\
5&127--103&  992 \\
\hline
\end{tabular}
\end{center}
\end{table}

\newpage

\begin{table}
\caption{Rank of lines on the basis of the power flux that should be extracted by the network ($\Delta P$) to restore a correct networks functioning when that line is removed. Line in second column, power flux to be extracted by the network (in MW) in third column.}
\begin{center}
\begin{tabular}{|c|c|c|}
\hline
rank& line & $\Delta P$ (MW) \\
\hline
1&214-184& 1532.4 \\
2&127-103& 1475.4\\
3&127-186& 954.0 \\
4&117-190& 916.3\\
5&117-165& 729.3 \\
\hline
\end{tabular}
\end{center}
\end{table}

\newpage

\begin{figure}
\begin{center}
\includegraphics[width=9 cm, angle=270]{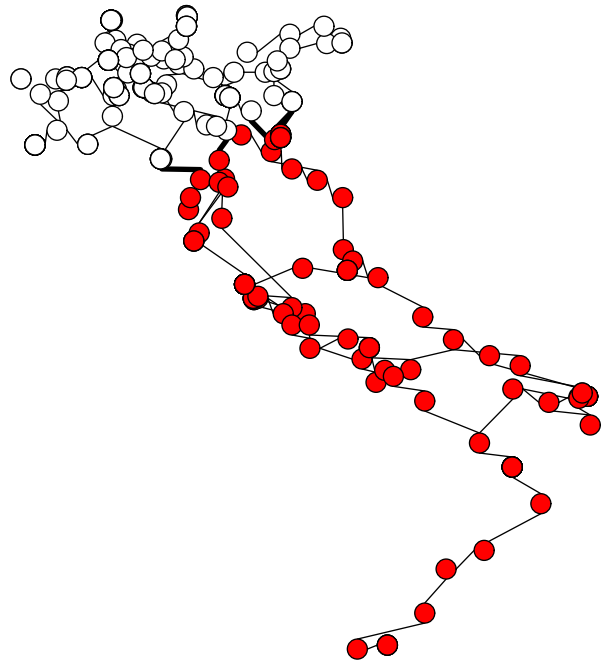}
\caption{Italian high-voltage (380 kV) transmission network (HVIET) resulting from the available data. Figure shows the first critical section (bold lines) resulting from the application of the min-cut theorem (see text). Nodes belonging to the two resulting subnetworks are highlighted in white and red.}
\end{center}
\label{f.1}
\end{figure}

\newpage

\begin{figure}
\includegraphics[width=6.0in]{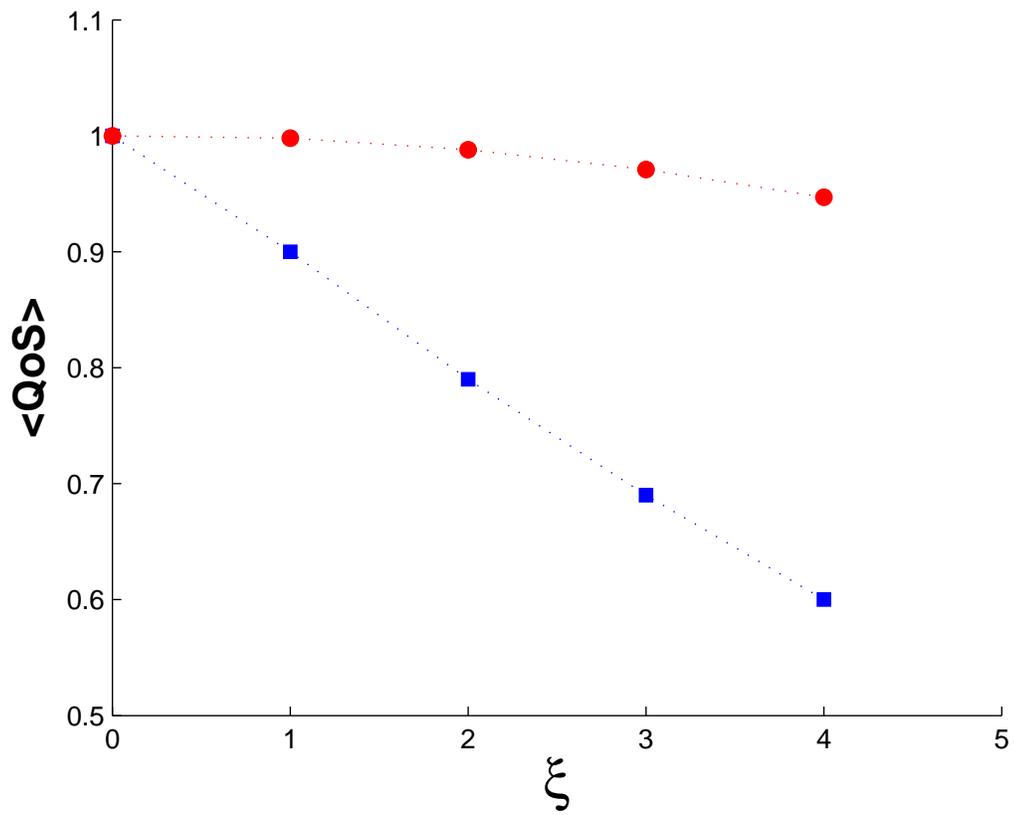}
\caption{Value of the $QoS$ as a function of the perturbation $\xi$. Squares represent $QoS$ values when no optimization is performed,  circles when re--dispatcing is performed.}
\end{figure}

\newpage

\begin{figure}
\includegraphics[width=6 in]{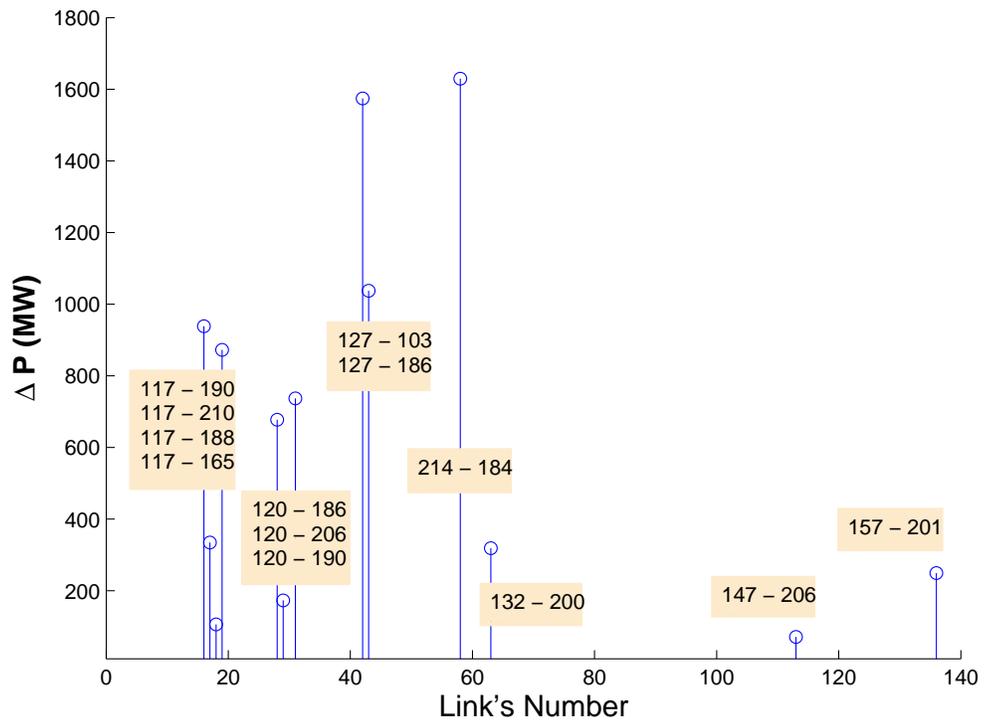}
\caption{Value of the quantity $\Delta P$ as a function of line's number. Lines whose removal produce the need of a larger re--dispatching are indicated.}
\end{figure}

\newpage

\begin{figure}
\begin{center}
\includegraphics[width=4.5 in,angle=270]{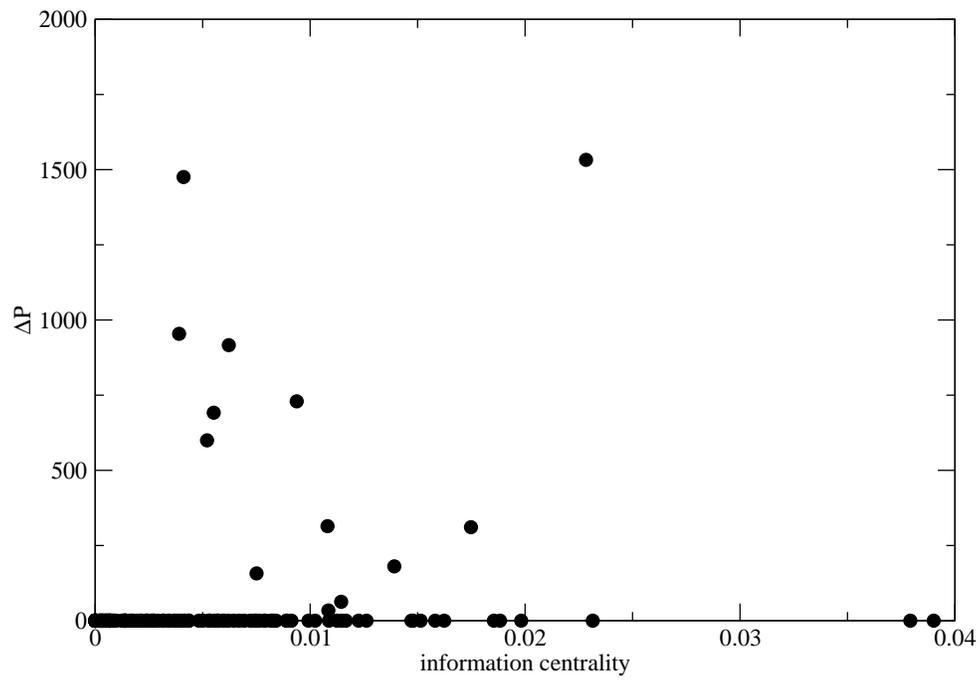}
\end{center}
\caption{Scatter plot relating $information$ centrality (x axis) and the value of $\Delta P$ (in MW), on the y-axis, of all the network's lines. Lines whose removal produces the disconnection of one (or more) nodes are not displayed in the figure as the $\Delta P$ value which is arbitrarily associated to them is set to $\infty$.}
\end{figure}

\end{document}